\documentclass[journal,twoside,web]{ieeecolor2}
\usepackage{generic}
\usepackage{cite}
\usepackage{amsmath,amssymb,amsfonts}
\usepackage{textcomp}
\usepackage{graphicx} 
\usepackage{hhline}
\usepackage{xcolor}
\usepackage{transparent}
\usepackage{url}
\usepackage{graphics}
\usepackage{subfigure}
\usepackage{booktabs}
\usepackage{siunitx}
\usepackage{multirow}
\usepackage{multicol}
\usepackage{cite}
\usepackage{color, soul}
\usepackage{comment}
\usepackage{tabularx}

\usepackage{algorithm}
\usepackage{algpseudocode}

\def\BibTeX{{\rm B\kern-.05em{\sc i\kern-.025em b}\kern-.08em
    T\kern-.1667em\lower.7ex\hbox{E}\kern-.125emX}}
\begin{document}
\title{Leveraging Self-Supervised Audio-Visual Pretrained Models to Improve Vocoded Speech Intelligibility in Cochlear Implant Simulation}
\author{Richard Lee Lai, Jen-Cheng Hou, I-Chun Chern, Kuo-Hsuan Hung, Yi-Ting Chen, Mandar Gogate, \\Tughrul Arslan, Amir Hussain, Chii-Wann Lin, and Yu Tsao, \IEEEmembership{Senior Member, IEEE}
\thanks{Richard Lee Lai, Jen-Cheng Hou,  I-Chun Chern, Yi-Ting Chen and Yu Tsao are with the Research Center for Information Technology Innovation, Academia Sinica, Taiwan.}
\thanks{Mandar Gogate and Amir Hussain are with
 the School of Computing, Edinburgh Napier University, Scotland, United Kingdom.}
\thanks{Tughrul Arslan is with the School of Computing, University of Edinburgh, United Kingdom.}
\thanks{Kuo-Hsuan Hung and Chii-Wann Lin are with the Department of Biomedical Engineering, National Taiwan University, Taiwan. corresponding e-mail: (cwlinx@ntu.edu.tw)} .
\thanks{* This work involved human subjects or animals in its research. Approval of all
ethical and experimental procedures and protocols was granted by Institutional
Review Board (IRB) on Biomedical Science Research, Academia Sinica under
Application No. AS-IRB-BM-21031, and performed in line with the “Study on the Effect of Speech Enhancement to Auditory Perception Under Normal Hearing and Cochlear Implant Simulations”.}
}

\maketitle

\begin{abstract}
\textbf{Objective:}
Individuals with hearing impairments face challenges in their ability to comprehend speech, particularly in noisy environments. This study explores the effectiveness of audio-visual speech enhancement (AVSE) in improving the intelligibility of vocoded speech in cochlear implant (CI) simulations.
\textbf{Methods:}
We propose a speech enhancement framework called Self-Supervised Learning-based AVSE (SSL-AVSE), which uses visual cues such as lip and mouth movements along with corresponding speech. Features are extracted using the AV-HuBERT model and refined through a bidirectional LSTM. Experiments were conducted using the Taiwan Mandarin speech with video (TMSV) dataset.
\textbf{Results:}
Objective evaluations showed improvements in PESQ from 1.43 to 1.67 and in STOI from 0.70 to 0.74. NCM scores increased by up to 87.2\% over the noisy baseline. Subjective listening tests further demonstrated maximum gains of 45.2\% in speech quality and 51.9\% in word intelligibility.
\textbf{Conclusion:}
SSL-AVSE consistently outperforms AOSE and conventional AVSE baselines. Listening tests with statistically significant results confirm its effectiveness. In addition to its strong performance, SSL-AVSE demonstrates cross-lingual generalization: although it was pretrained on English data, it performs effectively on Mandarin speech. This finding highlights the robustness of the features extracted by a pretrained foundation model and their applicability across languages.
\textbf{Significance:}
To the best of our knowledge, no prior work has explored the application of AVSE to CI simulations. This study provides the first evidence that incorporating visual information can significantly improve the intelligibility of vocoded speech in CI scenarios.

\end{abstract}


\section{Introduction}
Voice is essential for communication and psychological blending with society \cite{mccroskey1998communication}. The advancement of digital technologies has led to the emergence of various voice-related applications in the field of information and communications technology. According to the World Health Organization (WHO), one in four adults over 60 years of age and 15$\%$ of the general adult population experience hearing loss. Untreated hearing loss can lead to feelings of loneliness and result in isolation for the elderly while severely impairing learning ability in young children \cite{arlinger2003negative,pichora2013helping}. 
Research on hearing loss and the development of innovative techniques to support those affected has become a significant area of focus. According to the WHO’s classification of hearing impairment \cite{liao2011audiometer}, cochlear implants (CIs) are groundbreaking devices that restore hearing in individuals with severe-to-profound hearing loss \cite{zeng2008cochlear,clark2015multi,wilson2015getting,zeng2017challenges} and may also contribute to improved cognitive functioning \cite{moberly2019does,almomani2021cognitive}. CIs comprise an external sound processor and an internal component that delivers precisely timed electrical pulses to stimulate the auditory nerve. They have significantly improved the quality of life for hundreds of thousands of individuals with severe to profound hearing loss. Approved by the Food and Drug Administration (FDA) for individuals aged 12 months and older, CIs provide a highly effective means of restoring auditory perception.

Previous studies have confirmed that under quiet conditions, CI can effectively enhance the hearing capability of recipients, especially for speech recognition \cite{zeng2008cochlear, zeng2022celebrating, ghosh2023bilateral, ghosh2021cci}. However, it has been reported that speech recognition performance degraded considerably when the target speech signals are distorted \cite{chen2015evaluation, goehring2017speech, shekar2022convolutional, saba2022effects}. In real-world scenarios, there are several distortion sources, including background noise, reverberation, and interfering speech. To address speech distortion issues, a speech enhancement (SE) unit is usually adopted as a front-end processing unit in CI devices \cite{henry2021noise, wang2018speech}. Various techniques, including single-channel SE algorithms like spectral subtraction \cite{verschuur2006evaluation, kokkinakis2015evaluation}, subspace methods \cite{loizou2005subspace}, optimized gain functions \cite{mauger2012perceptually}, and commercial solutions \cite{dawson2011clinical,dingemanse2018optimising}, have all been applied to improve CI performance. Furthermore, multi-microphone and beamforming approaches have been explored for SE in CI users, taking advantage of spatial filtering to better isolate speech signals from background noise \cite{hersbach2013beamformer,kokkinakis2008using,kokkinakis2010multi, stronks2022beamforming}.

In recent years, SE techniques have improved significantly thanks to advances in machine learning algorithms. Notable examples include non-negative matrix factorization \cite{wilson2008speech, mohammadiha2013supervised}, sparse coding \cite{schmidt2007wind,sigg2010speech}, compressive sensing \cite{wang2016compressive}, and robust principal component analysis \cite{huang2012singing}. More recently, further advancements have been achieved through the powerful regression capabilities of deep learning–based models \cite{lu2013speech, xia2014wiener,  wang2018supervised, xu2015regression, zhang2016deep, erdogan2015phase, hsieh2020improving, weninger2015speech, yang2022improving, qi2019theory, qi2020tensor, kar2025improved, reddy2022performance, sowjanya2022mask, sivapatham2023deep, banduka2024delay, sivapatham2022gammatone}. For these approaches, deep neural networks are often used as a mapping function to carry out enhancement filtering on noisy input to attain high-quality speech signals. Several extensions have been made to these deep-learning models. One direction is to use a more suitable objective function to train the SE system. In \cite{kolbk2017speech, fu2018end, kolbaek2020loss, germain2018speech, fu2019metricgan, fu2021metricgan+, hsieh2020improving}, speech metric-oriented objective functions are derived, which can be divided into two categories. The first category directly considers a particular metric to form the objective function, such as \cite{kolbk2017speech, fu2018end, kolbaek2020loss, sivapatham2021performance}. The second uses another neural network model to form the objective function, such as \cite{germain2018speech, fu2019metricgan, fu2021metricgan+, hsieh2020improving}. Experimental results confirm that when a speech-metric oriented objective function is used, the SE system can be guided to achieve desirable output with optimal speech metric scores. In addition to designing more suitable objective functions, some researchers have attempted to incorporate information from other modalities as auxiliary inputs to the SE model, enabling exploitation of additional contextual information. Visual clues are one important modality that carries complementary information to speech signals during everyday communication. Numerous audio-visual multi-modal SE approaches, termed AVSE, have been proposed \cite{gabbay2018visual, ephrat2018looking, hou2018audio, michelsanti2019deep, iuzzolino2020av, sadeghi2020audio, michelsanti2021overview, chuang2022improved, chern2022audio, balasubramanian2023estimation}. These studies clearly show that visual cues can successfully enhance the performance of audio-only speech enhancement (AOSE).


Developing an efficient AVSE system with limited training data is a critical challenge in real-world applications. Following \cite{chern2022audio}, we address this issue by leveraging self-supervised learning (SSL) in AVSE. SSL models are trained by reconstructing the original input at the output, thereby learning to effectively analyze and resynthesize the data without requiring labels. Across numerous tasks, SSL has demonstrated its ability to extract more representative features, thereby boosting performance in downstream classification and regression tasks \cite{kolesnikov2019revisiting,doersch2015unsupervised}. For example, the well-known Bidirectional Encoder Representations from Transformers (BERT) model generates contextual language representations from large text corpora. As a versatile AI pretraining model originally developed for Natural Language Processing (NLP), BERT has demonstrated substantial performance improvements over previous supervised approaches across a wide range of tasks, including language understanding and speech recognition \cite{kenton2019bert,shin2019effective,huang2021speech}. In speech processing, HuBERT—a BERT-derived model based on hidden units \cite{hsu2021hubert}—has shown strong effectiveness for the SE task \cite{huang2022investigating, hung2022boosting}. Building on these advances, we propose a novel SSL-AVSE framework that leverages AV-HuBERT \cite{shi2022learning}, an audio-visual extension of HuBERT.

In the past decade, deep learning-based SE techniques have been applied to CI systems \cite{bolner2016speech,lai2016deep, lai2018deep, kang2021deep, mamun2024speech, balasubramanian2023ideal}, enabling more adaptive and robust processing methods tailored to the diverse auditory environments experienced by CI users. While the benefits of incorporating visual cues into the SE process are well-established, their specific advantages for CI devices have, to our knowledge, not yet been explored. In this study, we aim to evaluate the effectiveness of the proposed SSL-AVSE system for CI, with the implementation and user test scenario illustrated in Fig. 1. As shown in the figure, the SSL-AVSE system processes the combined noisy speech and video as inputs and outputs enhanced speech, which is then provided to a CI device. In this study, we employ a 16-channel speech vocoder to process enhanced utterances, simulating CI audio and calculating relevant performance metrics. Additionally, a listening test with 80 subjects is conducted to assess subjective speech quality and word intelligibility. Results indicate that the proposed SSL-AVSE model achieves maximum improvements of over 45\% in speech quality and 50\% in intelligibility compared with the original noisy signals, and over 25\% and 45\% improvements, respectively, compared with the baseline AVSE system. 


\begin{figure}
\centering
\includegraphics[width=0.48\textwidth]{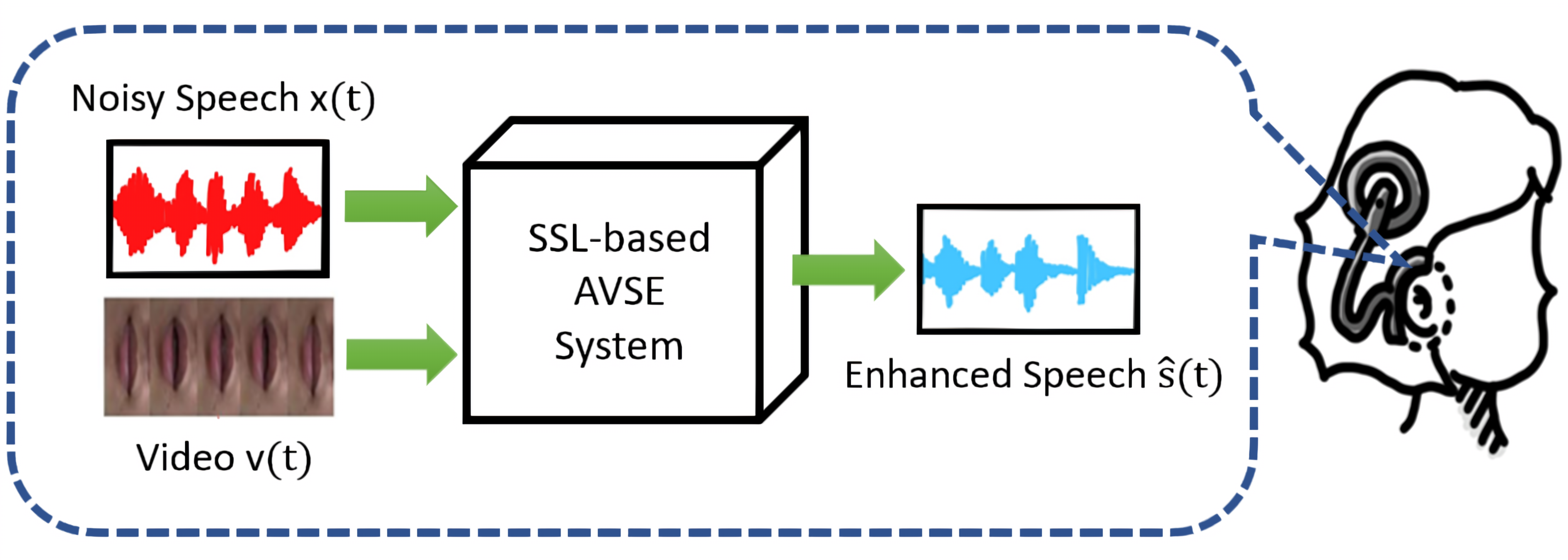}
\caption{Overview of our proposed system: Noisy speech and video are input into our SSL-based AVSE model, which outputs enhanced speech for CI users.}
\label{overview}
\end{figure}

\section{Methods}
In this section, we first formulate our problem and then present the SSL-AVSE network used in this study. We then demonstrate our training criteria and inference procedures, including the mathematical theories involved.

\subsection{Problem Formulation}

Given a speech signal $s(t)$ and a noise signal $n(t)$, the noisy signal $x(t)$ can then be denoted as: 

\begin{equation}
	x(t) = s(t) + n(t) \label{1}
\end{equation}

The noisy signal $x(t)$ is transformed into the spectral feature $X$. The model predicts a ratio mask $M$ to extract the clean speech signal for the corresponding target speaker from $X$. $M$ is estimated from our SSL-AVSE model, which uses the noisy signal $x(t)$ and additional visual cues $v(t)$ composed of the lip image sequence. The enhanced speech can then be obtained by the following formula:

\begin{equation}
	\hat{S} = X \otimes M \label{2}	
\end{equation}
where ``$\otimes$'' indicates an element-wise multiplication. From the above, we can then use the enhanced representation, $\hat{S}$, to reconstruct the waveform signal $\hat{s}(t)$, also known as enhanced speech.

\subsection{Audio-Visual Speech Enhancement Networks}
In this paper, we propose a novel SSL-AVSE framework, as illustrated in Fig.~\ref{SSL}. Specifically, SSL-AVSE integrates a Transformer-based AV-HuBERT network, a pretrained audio-visual foundation model, with the SE model.

\subsubsection{\bfseries Data Preprocessing}
 The model takes the visual stream of detected lip images from the target speaker $v(t)$, the noisy speech $x(t)$ as input, and outputs the enhanced speech $\hat{s}(t)$ for the target speaker while suppressing noise signals.
 
\begin{figure}

\centering
\includegraphics[width=0.48\textwidth]{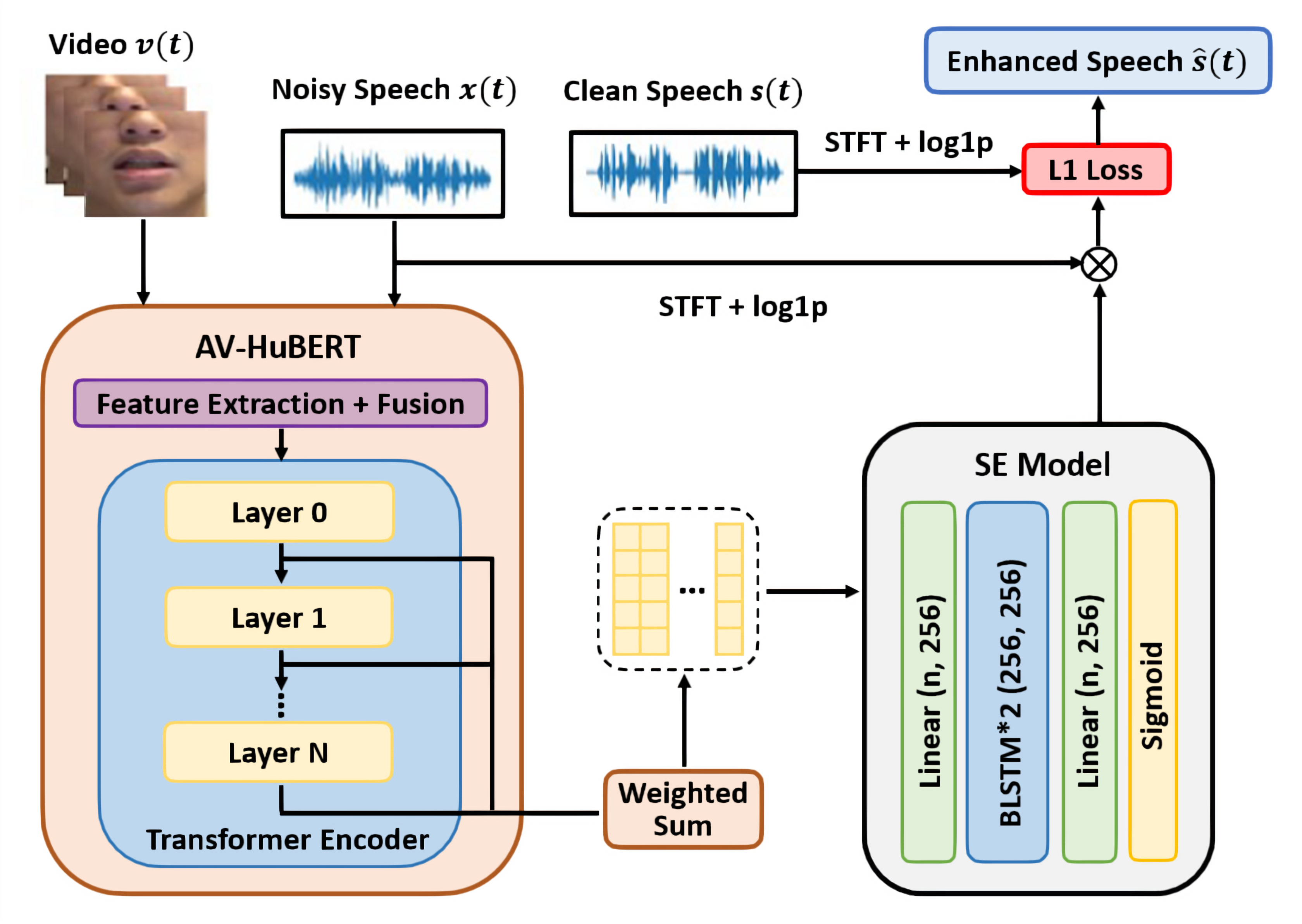}
\caption{The proposed SSL-AVSE model includes AV-HuBERT and SE modules. The AV-HuBERT model, consisting of multiple Transformer layers, is used to extract key features from noisy speech and lip images. An SE model then performs enhancement using them.}
\label{SSL}
\end{figure}

{\bfseries Audio Preprocessing.} In this study, the noisy speech signal $x(t)$ is first transformed into spectrograms using the Short-Time Fourier Transform (STFT) with an FFT size of 512, a window length of 400, and a hop size of 160. Subsequently, the $log1p$ function ($log1p(z)$ = log(1+$z$)) is applied to these spectrograms to extract $log1p$ spectral features. It has been demonstrated in our prior research that these $log1p$ spectral features outperform conventional log power spectral features in terms of SE performance \cite{hung2022boosting, fu12boosting}. As shown in Fig. \ref{SSL}, the noisy $log1p$ spectral features are multiplied by the estimated mask to produce enhanced speech. The mask is generated by the SSL-AVSE system, which is based on AV-Hubert and takes the video signal, $v(t)$, along with the noisy speech signal, $x(t)$,  as input. 

{\bfseries Video Preprocessing.} The video signal, $v(t)$, is comprised of sequential images sampled at 50 frames per second (fps), cropped around the target speaker's mouth region of interest (ROI). 
The ROI is detected using a facial landmark detector based on a two-dimensional Face Alignment Network (FAN) \cite{bulat2017far}, center-cropped to 88 × 88 pixels, and normalized using [0, 255] scaling followed by standardization with mean 0 and standard deviation 1.

\subsubsection{\bfseries Model} 
The representations of each Transformer-encoder layer are denoted as $H^l$, where $0\leq l\leq L-1$ and $L$ is the number of layers. A trainable function $w(\cdot)$ is then applied to all of the layer representations as follows:

\begin{equation}
H_{WS}=\sum_{l=0}^{L-1}w^lH^{l},\
\end{equation}
where $w^l$ is the weight of the $l$-th layer and has the properties $w^l\geq0$ and $\sum_{l}w^l=1$.

The extracted features are then passed to the SE model, which consists of a two-layer bidirectional long short-term memory (BLSTM) module positioned between two linear layers. The output of the SE module is a soft mask and is multiplied by the magnitude of the noisy speech spectra. This is then compared with the clean speech spectra to determine the L1 (absolute) loss.

While the video segment length used for learning SSL-AVSE is fixed, at the inference stage, our audio-visual extraction model can be applied to process videos of arbitrary length. This is done by applying a sliding window technique, which shifts the proposed window along the video segment until its entire length is covered. 



\subsection{CI Vocoded Speech}

We passed voice signals through a vocoder to simulate CI sounds. These simulations were then played to normal hearing (NH) people to conduct a listening test \cite{fetterman2002speech, shannon1995speech}. Compared with ordinary speech, vocoded speech is more difficult to understand by NH listeners due to the loss of spectral detail. Several studies have examined CI vocoder simulated speech on NH subjects in order to understand the associations between specific factors and CI users \cite{stickney2004cochlear,fu1998effects, friesen2001speech, lai2016deep}. Because accurate CI sounds are not always readily available, vocoder simulations can avoid the manifestation of patient-specific confounding factors, such as neural survival patterns \cite{loizou1999introduction}. Therefore, the CI vocoder can serve as an invaluable tool in related research. 

\begin{figure*}
\centering\includegraphics[width=0.9\textwidth]{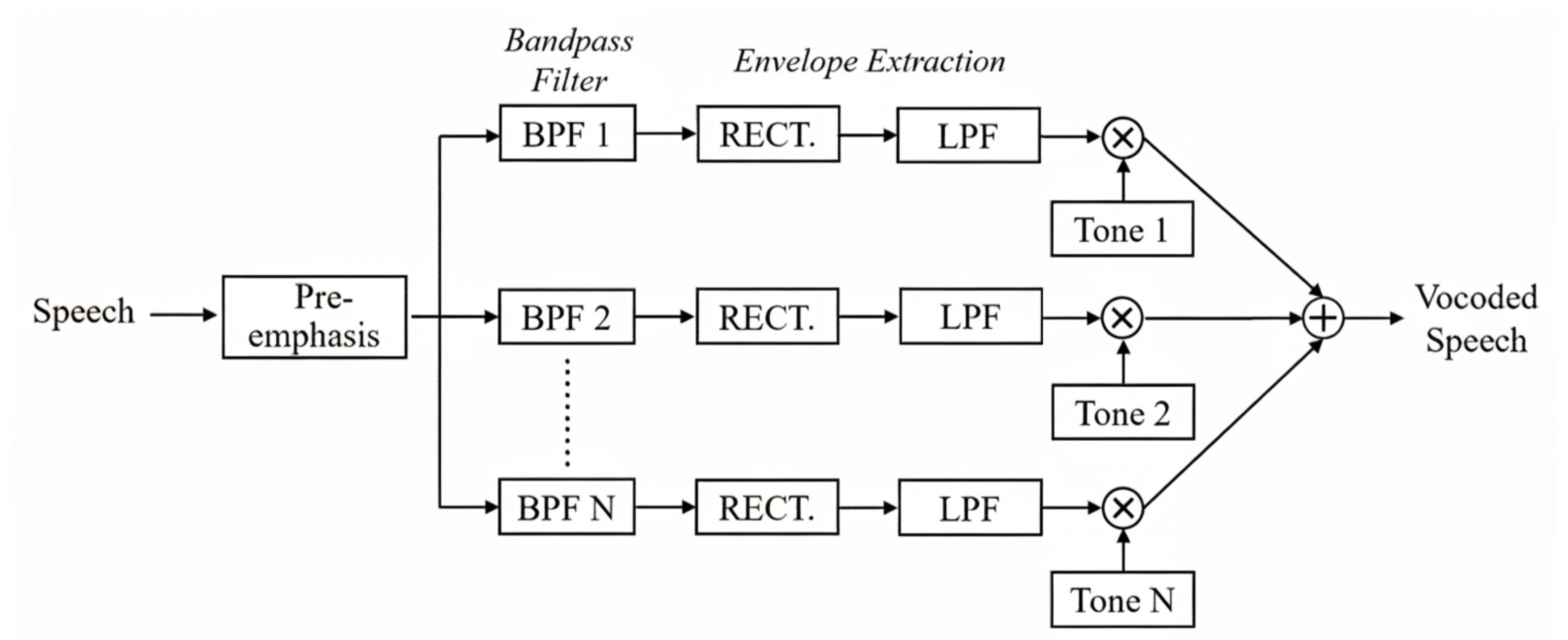}
\caption{Block diagram of the tone vocoder used in this study. The system consists of a set of band-pass filters, an envelope extractor, and sinewave carriers. The results of each band are then summed and combined to form vocoded speech.}
\label{tonevocoder}
\end{figure*}

To simulate CI audio in this study, a tone vocoder was used to process the speech signals following the procedure illustrated in Fig.~\ref{tonevocoder}. As shown in the figure, there are four steps: (1) 16 Butterworth band-pass filters were used to process an input temporal sequence to produce bandpass signals. (2) For each band waveform, a full-wave rectification function was leveraged to smooth the signal and to generate the corresponding envelope wave. (3) We added a tonal signal to the envelope to produce the modulated band voice. (4) We summed all modulated voice and performed a normalization operation to generate the vocoded speech, which has the identical root-mean-square value to the original input signal.

\subsection{Measures of Speech Quality and Intelligibility}

In this study, we employed four objective metrics to evaluate both non-vocoded and vocoded speech, namely: (1) for non-vocoded speech: the Perceptual Evaluation of Speech Quality (PESQ) \cite{rix2001perceptual}, the Short Time Objective Intelligibility (STOI) \cite{taal2011algorithm}, and the Levenshtein Phone Similarity (LPS) \cite{10363040}; for vocoded speech: the Normalized Covariance Metric (NCM) \cite{ma2009objective}. PESQ is a metric designed to predict subjective opinion scores of degraded speech samples, providing numerical values ranging from –0.5 to 4.5. To compute PESQ, the distorted speech sample must be paired with its corresponding clean reference signal, making PESQ an intrusive approach \cite{van2018evaluation}. STOI is a widely used measure of speech intelligibility in SE tasks. It has been shown to correlate well with the intelligibility of degraded speech signals and accounts for the effects of non-linear processing on noisy speech \cite{taal2011algorithm}. STOI scores range from 0 to 1. LPS measures the Levenshtein distance between phoneme sequences extracted from the enhanced and clean speech. Higher values denote stronger preservation of the intended phone sequence and are useful for detecting hallucinated \cite{saijo25_interspeech}.
NCM is a speech transmission index (TI)-related metric, estimated from the covariance of the envelopes between clean and processed signals. Prior studies have shown that NCM correlates well with the intelligibility of vocoded speech, largely because its calculation resembles CI processing strategies—both rely on envelope information across multiple frequency bands while discarding fine-structure details. NCM scores range from 0 to 1, with higher values indicating better intelligibility. Further details on the NCM measure can be found in \cite{ma2009objective}. To evaluate the effectiveness of the proposed SSL-AVSE, we used PESQ, STOI, and LPS to assess the SE results for non-vocoded speech. For CI-vocoded speech, performance was measured using NCM, as recommended in \cite{chen2011predicting}.

Subjective measures used in this study include overall speech quality and word intelligibility. The former was evaluated following the ITU-T P.808 protocol \cite{naderi2020open}, where subjects were asked to rate the quality of an entire utterance on a 5-point scale. The higher the score, the better the perceived quality of the recorded sentence. The latter is based on whether subjects could clearly understand individual words in the given sentence. Since each utterance contains a total of 10 Chinese characters, a score of 1.0 (= 10/10) indicates a complete understanding of each character in the given target sentence.


\subsection{Implementation Details}

\subsubsection{Ablation Studies}
To validate the effectiveness of incorporating a pretrained AV foundation model, we first built an AOSE system using a BLSTM network. We then added a ResNet-18 module to process visual cues and integrated it with the AOSE system to establish a baseline AVSE model.

\subsubsection{Model Training}
We optimized the SSL-AVSE model using partial fine-tuning (PF). In this setup, the convolutional layer weights in AV-HuBERT, used for feature extraction, are kept fixed, while the transformer encoder weights are fine-tuned from the pretrained checkpoint.
Training was conducted using the Adam optimizer with a weight decay of $10^{-4}$, a batchsize of 32, and an initial learning rate of $10^{-4}$. In addition, the learning rate is halved when errors are encountered. The proposed model was trained on the Taiwan Mandarin speech with video (TMSV) dataset \footnote{https://bio-asplab.citi.sinica.edu.tw/Opensource.html\#TMSV} for 50 epochs. We avoid overfitting on the training set by employing early-stopping techniques. 

\section{Experiments}
\subsection{Experimental Setup}
Noise can be roughly categorized into stationary and non-stationary based on the acoustic properties in the frequency domain. For example, monotonic background noises would fall under the category of stationary noise, while highly varied ones would be considered non-stationary. The removal of non-stationary noise is particularly challenging due to the substantial overlap and interference between speech and noise signals. To assess the effectiveness of our models, five non-stationary noise conditions were considered in the test set: ``babycry'' (sound of a baby crying), ``babble'' (multiple people talking simultaneously in a crowd), ``one talker'', ``two talkers'', and ``three talkers''. All these types are associated with human sounds or utterances. The amount of noise used to corrupt the input signal is measured by the signal-to-noise ratio (SNR) and expressed in decibels (dB).

The SE models were trained on the TMSV dataset, which includes video recordings of 18 native Mandarin speakers (13 male and 5 female), each contributing 320 utterances. Eight speakers (four male and four female) were used for training, while an additional unseen male speaker was reserved for testing. Each utterance consists of 10 Chinese characters and lasts between 2 to 4 seconds. The video was recorded with a resolution of 1,920 pixels $\times$ 1,080 pixels at 50 fps while the audio was recorded at a sampling rate of 48 \textit{K}Hz. The first 200 sentences were used to form the training set, while the last 120 sentences were used to form the test set.
To create noisy–clean speech pairs, the training utterances were artificially corrupted with 100 types of noise at five SNR levels ranging from –12 to 12 dB in 6 dB increments, yielding approximately 600 hours of noisy speech. Instead of using all generated utterances, the training set was constructed from 12,000 randomly selected noisy–clean pairs drawn from the full dataset. Testing was conducted on clean speech mixed with the five non-stationary noise types as mentioned earlier at SNRs ranging from -7 dB to 8 dB at increments of 3 dB. 

\subsection{Experimental Results}
\subsubsection{Effect of Fine-tuning the SSL Pretrained Model}
First, we present the speech quality (in terms of PESQ) and speech intelligibility (in terms of STOI) of SSL-AVSE with different pretrained AV-HuBERT models and verify the effectiveness of fine-tuning these models for AVSE. Table~\ref{ckpt_results} shows the PESQ and STOI scores of SSL-AVSE using several AV-HuBERT models pretrained on different datasets, including LRS3~\cite{afouras2018lrs3} and VoxCeleb2~\cite{chung18b_interspeech}, and noise augmentation, provided by the authors in~\cite{hsu2021hubert}. From the table, we can observe that fine-tuning a pretrained AV-HuBERT model with more diverse data leads to enhanced results. Since both LRS3 and VoxCeleb2 are English datasets, while the testing data consists of Mandarin speech, the results in Table~\ref{ckpt_results} demonstrate the potential of the proposed SSL-AVSE method to leverage high-resource languages for applications in low-resource languages.  Furthermore, the results in Table~\ref{ckpt_results} confirm the effectiveness of fine-tuning AV-HuBERT within our method. In the following experiments, SSL-AVSE denotes the one fine-tuning the AV-HuBERT pretrained on LRS3, VoxCeleb2, and noise augmentation, which is the best setup in Table~\ref{ckpt_results}.

\begin{table}
\caption{Objective scores of the speech enhanced by fine-tuning pretrained AV-HuBERT models. L, V, and N represent LRS3, VoxCeleb2, and noise augmentation, respectively, and w/o denotes without fine-tuning the AV-HuBERT model.}
\centering
\begin{tabularx}{0.9\columnwidth}{|>{\centering}m{2.25cm}|>{\centering\arraybackslash}X|>{\centering\arraybackslash}X|>{\centering\arraybackslash}X|>{\centering\arraybackslash}X|}
\hline

&\textbf{PESQ}&\textbf{STOI}\\
\hline
\textbf{Noisy} & 1.434 & 0.695  \\
\hline
\textbf{SSL-AVSE (L/V/N. w/o fine-tuning)} & 1.565 & 0.719  \\
\hline
\textbf{SSL-AVSE (L)} & 1.615 & 0.728  \\
\hline
\textbf{SSL-AVSE (L/V)} & 1.651 & 0.737  \\
\hline
\textbf{SSL-AVSE (L/V/N)} & \textbf{1.665} & \textbf{0.738} \\
\hline
\end{tabularx}
\label{ckpt_results}
\end{table}

\subsubsection{Objective Results}

\begin{table*}

\caption{Objective PESQ scores for non-vocoded speech. We can see that the lower the SNR, the greater the difference between the PESQ scores of SSL-AVSE-enhanced speech and those enhanced by either AVSE or AOSE. At higher SNRs, the differences between enhancement methods become less pronounced. "A" stands for audio-only, while 
"AV" stands for audio-visual.}

\centering
\begin{tabularx}{1.5\columnwidth}{|>{\centering}m{1.5cm}|>{\centering\arraybackslash}X|>{\centering\arraybackslash}X|>{\centering\arraybackslash}X|>{\centering\arraybackslash}X|>{\centering\arraybackslash}X|>{\centering\arraybackslash}X|>{\centering\arraybackslash}X|}
\hline

&\textbf{modality} &\textbf{-7dB}  &\textbf{-4dB} &\textbf{-1dB} &\textbf{2dB} &\textbf{5dB} &\textbf{8dB} \\
\hline
\textbf{Noisy} & N/A & 1.214 & 1.260 & 1.344 & 1.449 & 1.586 & 1.771 \\
\hline
\textbf{AOSE} & A & 1.227 & 1.296 & 1.403 & 1.540 & 1.713 & 1.934 \\
\hline
\textbf{ConvTasNet \cite{Luo_2019}} & A & 1.242 & 1.311 & 1.424 & 1.569 & 1.764 & 2.001 \\
\hline
\textbf{VisualVoice \cite{gao2021visualvoice}} & AV & \textbf{1.283} & 1.349 & 1.416 & 1.486 & 1.667 & 1.792 \\
\hline
\textbf{AVSE-VAE \cite{sadeghi2021mixture}} & AV & 1.217 & 1.285 & 1.388 & 1.511 & 1.677 & 1.890 \\
\hline
\textbf{AVSE} & AV & 1.226 & 1.324 & 1.453 & 1.591 & 1.773 & 2.002 \\
\hline
\textbf{SSL-AVSE} & AV & 1.271 & \textbf{1.353} & \textbf{1.474} & \textbf{1.619} & \textbf{1.801} & \textbf{2.020} \\
\hline

\end{tabularx}

\label{pesq}
\end{table*}

\begin{table*}
\caption{Objective STOI scores for non-vocoded speech. We can see that the lower the SNR, the greater the difference between the STOI scores of SSL-AVSE-enhanced speech and those enhanced by either AVSE or AOSE. For higher SNRs, the results for different enhancement methods converge. "A" stands for audio-only, while "AV" stands for audio-visual.}

\centering
\begin{tabularx}{1.5\columnwidth}{|>{\centering}m{1.5cm}|>{\centering\arraybackslash}X|>{\centering\arraybackslash}X|>{\centering\arraybackslash}X|>{\centering\arraybackslash}X|>{\centering\arraybackslash}X|>{\centering\arraybackslash}X|>{\centering\arraybackslash}X|}
\hline

&\textbf{modality} &\textbf{-7dB} &\textbf{-4dB} &\textbf{-1dB} &\textbf{2dB} &\textbf{5dB} &\textbf{8dB} \\
\hline
\textbf{Noisy} & N/A & 0.536 & 0.598 & 0.644 & 0.731 & 0.794 & 0.848 \\
\hline
\textbf{AOSE} & A & 0.523 & 0.590 & 0.663 & 0.733 & 0.797 & 0.854 \\
\hline
\textbf{ConvTasNet \cite{Luo_2019}} & A & 0.549 & 0.611 & 0.682 & \textbf{0.753} & 0.808 & 0.859 \\
\hline
\textbf{VisualVoice \cite{gao2021visualvoice}} & AV & \textbf{0.591} & 0.635 & 0.687 & 0.729 & 0.767 & 0.810 \\
\hline
\textbf{AVSE-VAE \cite{sadeghi2021mixture}} & AV & 0.516 & 0.580 & 0.647 & 0.714 & 0.774 & 0.827 \\
\hline
\textbf{AVSE} & AV & 0.549 & 0.616 & 0.685 & 0.751 & \textbf{0.809} & \textbf{0.860}  \\
\hline
\textbf{SSL-AVSE} & AV & 0.582 & \textbf{0.636} & \textbf{0.695} & \textbf{0.753} & \textbf{0.809} & 0.858 \\
\hline

\end{tabularx}

\label{stoi}
\end{table*}

\begin{table*}
\caption{Objective LPS scores for non-vocoded speech. We can see that the lower the SNR, the greater the difference between the LPS scores of SSL-AVSE-enhanced speech and those enhanced by either AVSE or AOSE. At higher SNRs, the differences between enhancement methods become less pronounced. "A" stands for audio-only, while "AV" stands for audio-visual.}

\centering
\begin{tabularx}{1.5\columnwidth}{|>{\centering}m{1.5cm}|>{\centering\arraybackslash}X|>{\centering\arraybackslash}X|>{\centering\arraybackslash}X|>{\centering\arraybackslash}X|>{\centering\arraybackslash}X|>{\centering\arraybackslash}X|>{\centering\arraybackslash}X|}
\hline

&\textbf{modality} &\textbf{-7dB} &\textbf{-4dB} &\textbf{-1dB} &\textbf{2dB} &\textbf{5dB} &\textbf{8dB} \\
\hline
\textbf{Noisy} & N/A & 0.146 & 0.187 & 0.244 & 0.322 & 0.417 & 0.528  \\
\hline
\textbf{AOSE} & A & 0.196  & 0.235 & 0.281 & 0.349 & 0.444 & 0.555   \\
\hline
\textbf{AVSE-VAE \cite{sadeghi2021mixture}} & AV & 0.167 & 0.197 & 0.246 & 0.322 & 0.405 & 0.495   \\
\hline
\textbf{AVSE} & AV & 0.204 & 0.254 & 0.311 & 0.38 & 0.483 & 0.587   \\
\hline
\textbf{SSL-AVSE} & AV & \textbf{0.275} & \textbf{0.32} & \textbf{0.387} & \textbf{0.463} & \textbf{0.559} & \textbf{0.639} \\
\hline

\end{tabularx}

\label{phone}
\end{table*}

\begin{table}
\caption{Objective NCM scores for vocoded speech. Similar to the results of STOI scores, We can see that the lower the SNR, the greater the difference between the NCM scores of SSL-AVSE-enhanced speech and those enhanced by either AVSE or AOSE. However, unlike the STOI results, there is no marked convergence between the absolute amount of improvement between the NCM scores of the noisy baseline and that of utterances enhanced by SSL-AVSE.}

\centering
\begin{tabularx}{0.95\columnwidth}{|>{\centering}m{1.5cm}|>{\centering\arraybackslash}X|>{\centering\arraybackslash}X|>{\centering\arraybackslash}X|>{\centering\arraybackslash}X|>{\centering\arraybackslash}X|>{\centering\arraybackslash}X|>{\centering\arraybackslash}X|}
\hline

&\textbf{-7dB}&\textbf{-4dB}&\textbf{-1dB} &\textbf{2dB} &\textbf{5dB} &\textbf{8dB}\\
\hline
\textbf{Noisy} & 0.226 & 0.321 & 0.416 & 0.519 & 0.600 & 0.671 \\
\hline
\textbf{AOSE} & 0.354 & 0.441 & 0.553 & 0.630 & 0.724 & 0.811 \\
\hline
\textbf{AVSE} & 0.387 & 0.482 & 0.581 & 0.681 & 0.773 & \textbf{0.849} \\
\hline
\textbf{SSL-AVSE} & \textbf{0.423} & \textbf{0.507} & \textbf{0.598} & \textbf{0.690} & \textbf{0.777} & 0.846 \\
\hline

\end{tabularx}

\label{ncm}
\end{table}

As shown in Tables~\ref{pesq}–\ref{ncm}, the proposed SSL-AVSE consistently outperformed the baseline AOSE and AVSE systems, achieving higher objective measures of speech quality, intelligibility, and LPS across most noise conditions. The difference was particularly notable for low SNRs. For an SNR of -7 dB, the PESQ, STOI, LPS and NCM values of SSL-AVSE were 3.7\% (from 1.226 to 1.271), 6.0\% (from 0.549 to 0.582), 34.8\% (from 0.204 to 0.275), and 9.3\% (from 0.387 to 0.423) higher than those of AVSE, respectively, while they were 3.6\% (from 1.227 to 1.271), 11.3\% (from 0.523 to 0.582), 40.3\% (from 0.196 to 0.275), and 19.5\% (from 0.354 to 0.423) higher than those of AOSE, respectively. The results were even more striking when compared with the noisy baseline; PESQ, STOI, LPS, and NCM values increased by 4.7\% (from 1.214 to 1.271), 8.6\% (from 0.536 to 0.582), 88.4\% (from 0.146 to 0.275), and 87.2\% (from 0.226 to 0.423), respectively. At higher SNRs, the differences in PESQ, STOI, and NCM scores between SSL-AVSE and AVSE are less pronounced, indicating that the pretrained AV-HuBERT model provides greater benefits for SE under challenging conditions. The results also show smaller improvements across all three metrics for speech enhanced by SSL-AVSE compared to AOSE at higher SNR levels, indicating that visual cues provide relatively less benefit for AVSE performance under these conditions.

\begin{table*}
\caption{Paired t-tests of speech quality results were used to compare AOSE, AVSE, and SSL-AVSE against the noisy baseline for each specific noise type and SNR condition. The degree of freedom for all values is 19. A p-value of less than 0.05 implies statistical significance.}

\centering
\begin{tabular}{|l|c|c|c|c|c|c|c|c|}
\hline
 & \multicolumn{2}{|c|}{\textbf{babble 2 dB}} & \multicolumn{2}{|c|}{\textbf{babble 5 dB}} & 
 \multicolumn{2}{|c|}{\textbf{babycry 2 dB}} &
  \multicolumn{2}{|c|}{\textbf{babycry 5 dB}}
\\  
\hline
\textbf{} & \textbf{t-value} & \textbf{$p$-value} & \textbf{t-value} & \textbf{$p$-value} & \textbf{t-value} & \textbf{$p$-value} & \textbf{t-value} & \textbf{$p$-value}\\
\hline
\textbf{AOSE} & 1.571 & $<$0.05 & 1.746 & $<$0.05 & 1.191 & $<$0.05 & 3.145 &$<$0.05\\
\hline
\textbf{AVSE} & 1.337 & $<$0.01 & 0.883 & $<$0.01 & 1.047 & $<$0.01 & 2.732 &$<$0.05 \\
\hline
\textbf{SSL-AVSE} & 2.743 & $<$0.001 & 1.402 & $<$0.001 & 1.031 & $<$0.001 & 1.433 & $<$0.01\\
\hline
\end{tabular}
\label{quality}
\end{table*}

\begin{table*}
\caption{Paired t-tests of speech intelligibility results were used to compare AOSE, AVSE, and SSL-AVSE against the noisy baseline for each specific noise type and SNR condition. The degree of freedom for all values is 19. A p-value of less than 0.05 implies statistical significance.}

\centering
\begin{tabular}{|l|c|c|c|c|c|c|c|c|}
\hline
 & \multicolumn{2}{|c|}{\textbf{babble 2 dB}} & \multicolumn{2}{|c|}{\textbf{babble 5 dB}} & 
 \multicolumn{2}{|c|}{\textbf{babycry 2 dB}} &
  \multicolumn{2}{|c|}{\textbf{babycry 5 dB}}
\\  
\hline
\textbf{} & \textbf{t-value} & \textbf{$p$-value} & \textbf{t-value} & \textbf{$p$-value} & \textbf{t-value} & \textbf{$p$-value} & \textbf{t-value} & \textbf{$p$-value}\\
\hline
\textbf{AOSE} & 2.518 & $<$0.05 & 3.452 & $<$0.05 & 2.346 & $<$0.05 & 2.128 &$<$0.05\\
\hline
\textbf{AVSE} & 1.664 & $<$0.01 & 3.683 & $<$0.05 & 2.038 & $<$0.05 & 3.038 &$<$0.01 \\
\hline
\textbf{SSL-AVSE} & 2.842 & $<$0.001 & 4.535 & $<$0.01 & 1.406 & $<$0.01 & 4.406 & $<$0.01\\
\hline
\end{tabular}
\label{intelligibility}
\end{table*}

\subsubsection{Spectrogram Analysis}
\begin{figure}
\includegraphics[width=0.5\textwidth]{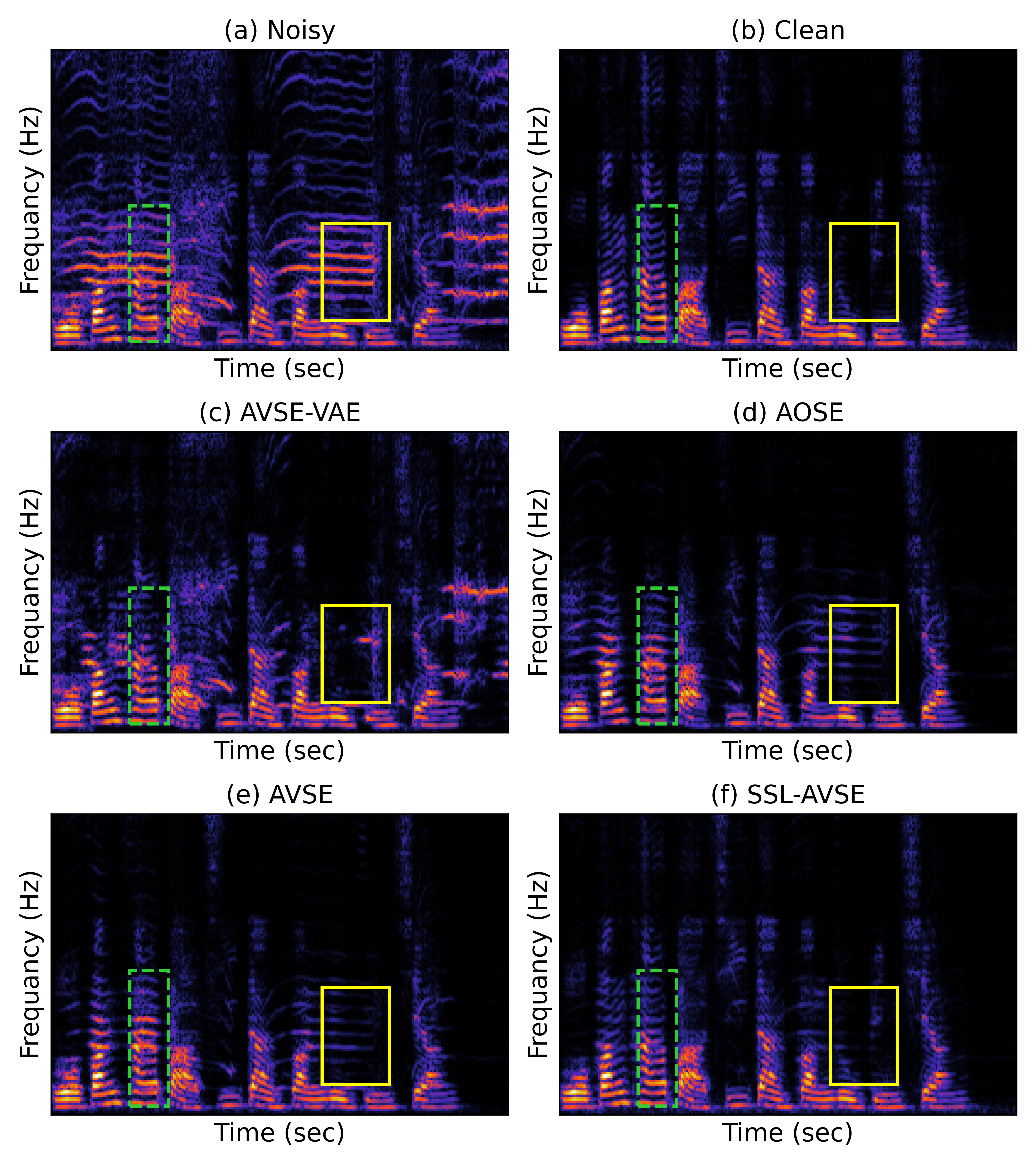}
\caption{Spectrograms of noisy, clean, AVSE-VAE enhanced, AOSE enhanced, AVSE enhanced, and SSL-AVSE-enhanced speech signals for ``babycry 2 dB'' noise. Note that SSL-AVSE enhanced speech preserves speech structures within the range of human speech more than those of other enhancement methods.}
\label{spec1}
\end{figure}

\begin{figure}
\includegraphics[width=0.48\textwidth]{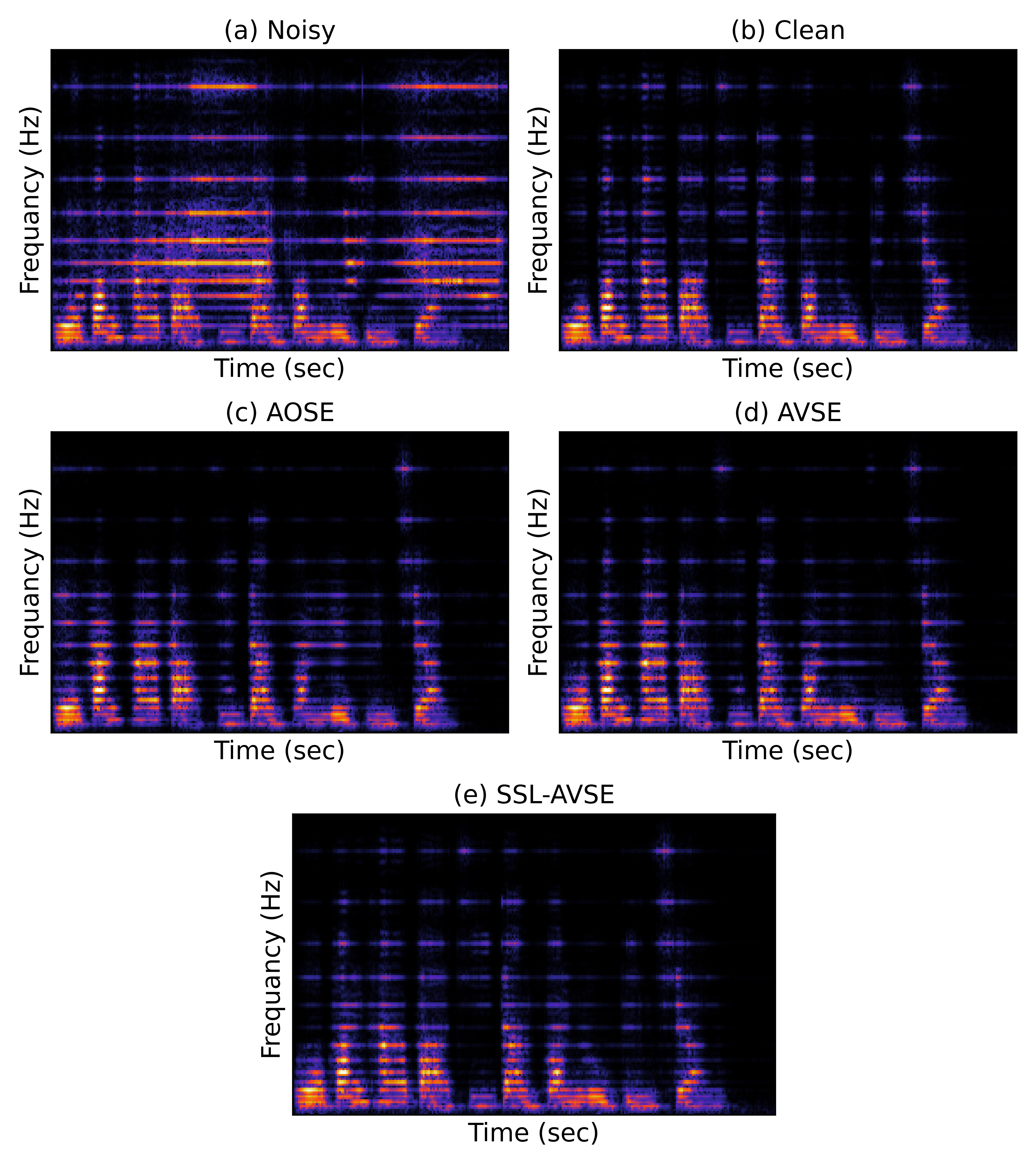}
\caption{Spectrograms of vocoded speech for noisy, clean, AOSE enhanced, AVSE enhanced, and SSL-AVSE-enhanced speech signals for ``babycry 2 dB'' noise. Note that SSL-AVSE enhanced speech has clearer structures compared with those of other enhancement methods.}
\label{spec2}
\end{figure}

A spectrogram plot is frequently employed to visually represent the time–frequency characteristics of a speech signal. In Fig.~\ref{spec1}, we present spectrograms of a noisy speech signal at a 2 dB SNR,  enhanced using four methods: AVSE-VAE \cite{sadeghi2021mixture}, AOSE, AVSE, and SSL-AVSE. Additionally, the spectra of the corresponding clean speech are included for comparison. Two regions of interest are highlighted in the spectrograms: a noise-only segment (yellow box) and a mixed speech–noise region (green dashed box). In the noise-only regions (yellow box), SSL-AVSE shows significant improvements over the baseline methods by effectively reducing noise. In the mixed speech–noise regions (green dashed box), SSL-AVSE introduces fewer distortions in the reconstructed speech compared with the other approaches.
In Fig.~\ref{spec2}, we showcase spectra of vocoded speech. From the figure, it is evident that the spectra of vocoded speech processed by SSL-AVSE preserve much clearer speech structures compared with AOSE and AVSE.

\subsubsection{Subjective Results}
We conducted listening tests under four conditions, comprising two noise types (``babycry" and ``babble") and two SNRs (2 dB and 5 dB). Each condition contained 120 utterances, with 20 drawn from each of six categories:
AOSE, AVSE, SSL-AVSE, clean, logarithm minimum mean squared error (logMMSE), and noisy speech. The logMMSE is a traditional SE method that enhances speech by minimizing the mean-square error in the logarithmic spectral domain \cite{ephraim1985speech}. Similar to \cite{lai2016deep}, we also include logMMSE as part of our benchmark.
Since our objective is to maximize SE in CI devices, we conducted our experiment using vocoded speech. A total of 80 participants (mean age: 35.7 years) took part in the study, with 20 individuals assigned to each noise type at a specific SNR condition to minimize cross-referencing bias. The listening tests were conducted in a quiet room, while the utterances were uploaded onto a listening test system that presented them randomly to the users. 

\begin{figure}
\includegraphics[width=0.5\textwidth]{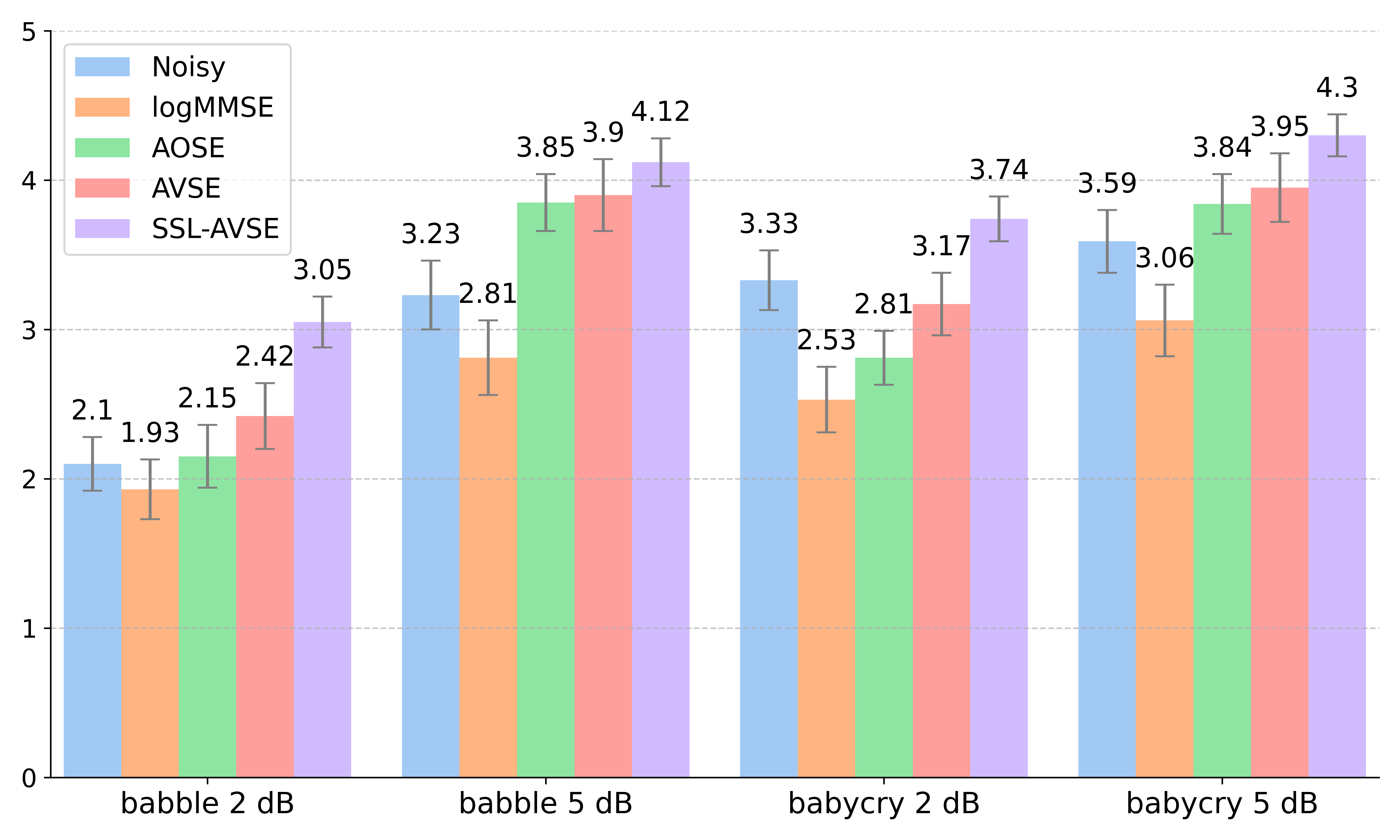}
\caption{Subjective speech quality scores for vocoded speech enhanced with different models. The x-axis represents the noise type, while the y-axis represents the speech quality score. For all noise types, SSL-AVSE is shown to perform notably better than other methods, with the greatest improvements occurring for noises with an SNR of 2 dB.}
\label{sound_quality}
\end{figure}

\begin{figure}
\includegraphics[width=0.5\textwidth]{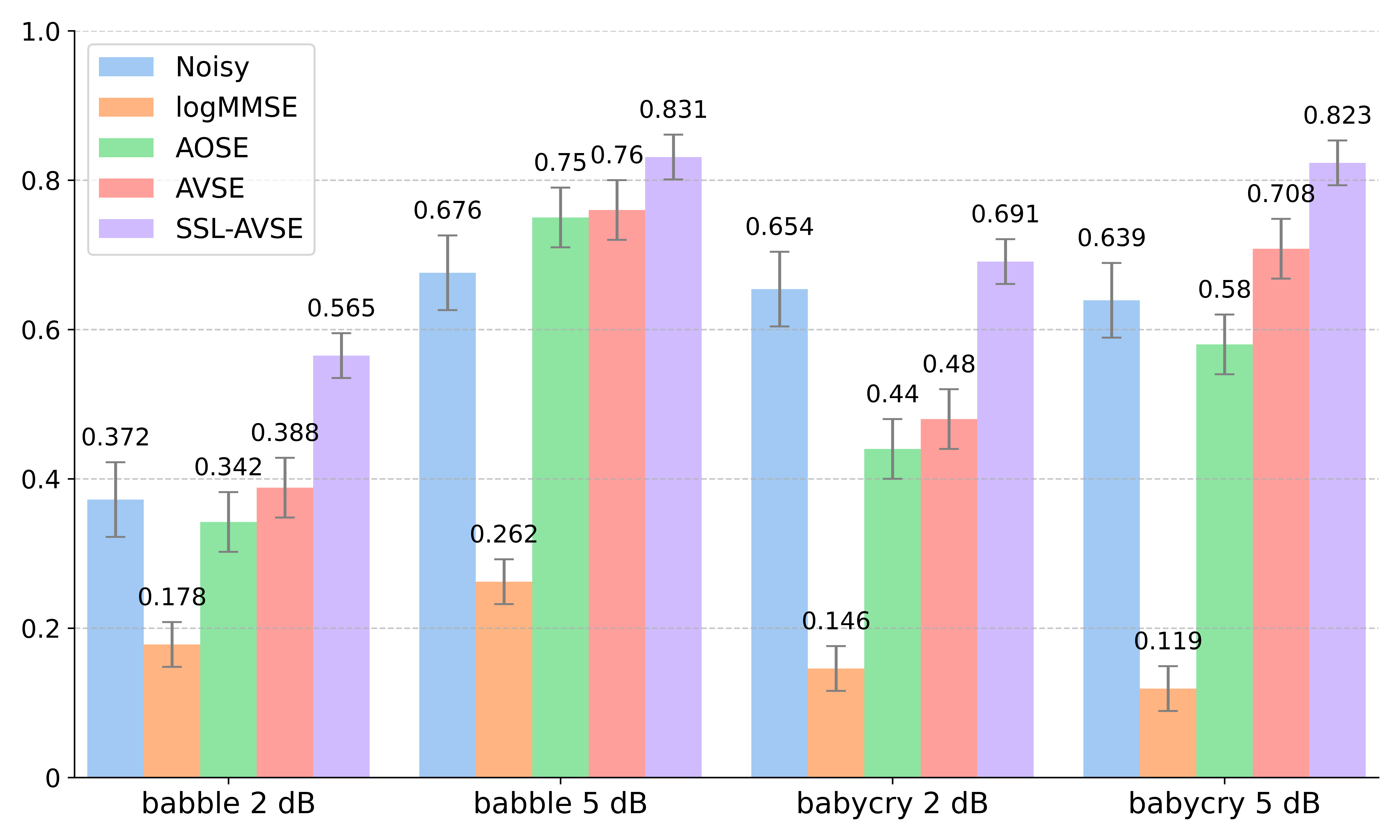}
\caption{Subjective character intelligibility scores of speech enhanced using different models. The x-axis represents the noise type, while the y-axis represents the word intelligibility score. As in the case of speech quality, SSL-AVSE is also shown to perform notably better than other methods, with the greatest improvements occurring for noises with an SNR of 2 dB.}
\label{word_intelligibility}
\end{figure}

As shown in Fig.~\ref{sound_quality} and Fig.~\ref{word_intelligibility}, results show that for both speech quality and word intelligibility, SSL-AVSE outperformed both AVSE and AOSE. For speech quality scores, when subjected to the most challenging ``babble 2 dB'' noise condition, the SSL-AVSE model exhibited a 26.0\% increase (from 2.42 to 3.05) for the former and a 41.9\% increase (from 2.15 to 3.05) for the latter, respectively. For word intelligibility scores, the improvement over the former and the latter were 45.6\% (from 0.388 to 0.565) and 65.2\% (from 0.342 to 0.565), respectively.

\begin{figure*}
\includegraphics[width=0.97\textwidth]{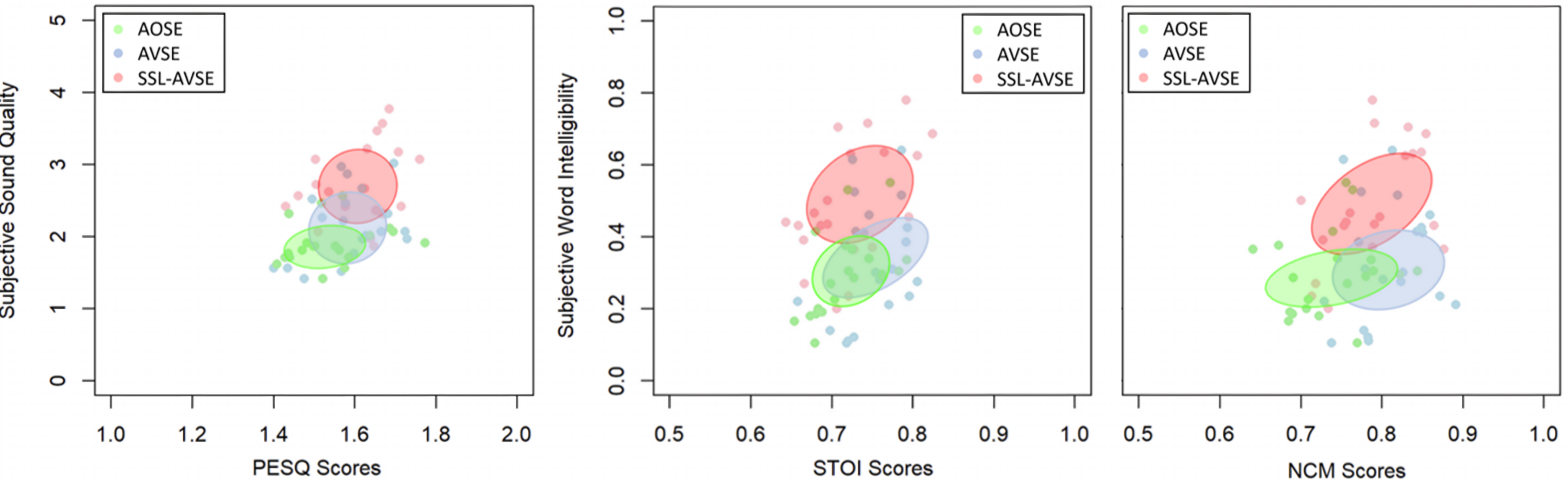}
\caption{Relationship between the subjective and objective scores of speech enhanced using different models. We choose utterances corrupted with ``babble 2 dB'' noise. The y-axis represents the subjective score, while the x-axis represents the objective score. The centers of the ovals represent the mean objective and subjective scores. There is clear segregation between utterances enhanced using the three different methods, with SSL-AVSE having the best effect.}
\label{ncm_intelligibility}
\end{figure*}
We further conducted paired t-tests on the quality and intelligibility results of listening tests under SSL-AVSE, AVSE, and AOSE conditions to verify their statistical significance. Results shown in Tables~\ref{quality} and~\ref{intelligibility} indicate that speech enhanced by all three methods differ statistically from noisy speech, with $p$-values all less than 0.001 for the ``babble'' noise and $p$-values all less than 0.05 for the ``babycry'' noise. Since the ``babble'' noise condition results in smaller $p$-values than those of ``babycry'', this demonstrates that our model actually performs better under more challenging noise conditions, especially those with multiple talkers. Further comparisons between speech enhanced by SSL-AVSE and that enhanced by AVSE or AOSE revealed significant statistical differences, reinforcing the promising capability of the proposed method.

\subsubsection{Comparison between Objective and Subjective Results}

Both objective and subjective evaluations demonstrated that SSL-AVSE yielded improved performance across different SNR levels. Moreover, speech signals with lower SNRs exhibited greater improvements in quality and intelligibility, underscoring the strong capability of SSL-AVSE in challenging noisy environments. The subjective results further validate the effectiveness of our model for vocoded speech, showing notable improvements compared with the noisy baseline. To more explicitly explore the relationship between objective and subjective results, three panels are presented in Fig. \ref{ncm_intelligibility}, focusing on utterances corrupted with ``babble 2 dB'' noise. For all three panels, the x-axis represents the subjective score, while the y-axis represents the objective score. Analyzing the left panel of Fig. \ref{ncm_intelligibility}, it is evident that SSL-AVSE outperforms AVSE and AOSE notably in terms of both objective and subjective speech quality scores, as indicated by the score distributions located towards the top-right side. Furthermore, examining the center and right panels of Fig.~\ref{ncm_intelligibility}, it is evident that SSL-AVSE achieves the best performance for both objective and subjective speech intelligibility scores, as indicated by the score distributions also located towards the top-right side.

\section{Discussion}

In this paper, we propose a novel SSL-AVSE approach that leverages AV-HuBERT and evaluate its effectiveness in improving speech perception in CI simulations. As shown in Table~\ref{ncm}, SSL-AVSE yielded higher NCM scores, with relative improvements ranging from 26.1$\%$ (0.671 to 0.846 at 8 dB) to 87.2$\%$ (0.226 to 0.423 at –7 dB) compared with the noisy baseline. This demonstrates the effectiveness of our model on vocoded speech signals. From Figs.~\ref{sound_quality} and~\ref{word_intelligibility},  subjective results also showed improvements of between 19.8\% and 45.2\% for speech quality (from 3.59 to 4.30 for ``babycry 5dB'' and from 2.10 to 3.05 for ``babble 2dB'') and improvements of between 5.7\% and 51.9\% for word intelligibility (from 0.654 to 0.691 for ``babycry 2dB'' and from 0.372 to 0.565 for ``babble 2dB'') when compared with the noisy baseline, further confirming the effectiveness of the proposed SSL-AVSE approach.

Our experimental results also indicate that the proposed SSL-AVSE model, although leveraging an audio-visual foundation model pretrained on English corpora, can still achieve strong performance on Mandarin AVSE tasks. We believe that this cross-lingual transferability is partly due to two factors. First, regression tasks such as speech enhancement and separation tend to be less language-dependent because they primarily focus on improving acoustic properties rather than modeling linguistic content. Prior work has shown that models trained exclusively on English can generalize well to other languages (e.g., Spanish and German) for SE without additional adaptation \cite{zhang19e_interspeech}. Second, audio-visual alignment in models like AV-HuBERT is largely based on shared phoneme-level features, which are relatively consistent across languages. This allows the learned correlations between phonemes and visual articulations (e.g., lip shapes) to transfer more effectively. Recent studies \cite{chern2022audio, dashtipour24_avsec} have shown that English-pretrained models (e.g., AV-HuBERT, U-Net) can improve Mandarin AVSE performance even without retraining the visual front-end, whereas word-level tasks such as lip reading often require language-specific mechanisms \cite{kim2023lip}. These observations suggest why the SSL-AVSE system can generalize effectively to Mandarin tasks despite being pretrained on English corpora.

\begin{table}[t]
\caption{Model complexity and inference latency of the proposed SSL-AVSE system.}
\centering
\begin{tabular}{|>{\raggedright\arraybackslash}p{2cm}|>{\raggedright\arraybackslash}p{5cm}|} 
\hline
 \textbf{Metric} & \textbf{Value} \\
 \hline
 \textbf{Parameters} & 106 M (103 M for AV-HuBERT) \\
 \hline
 \textbf{FLOPs} & 27.5 GFLOPs \\
 \hline
 \textbf{Model Size} & 423 MB (FP32) / 107 MB (INT8) \\
 \hline
 \multirow{2}{*}{\textbf{Latency}} 
 & -- 5 ms per 10 ms audio frame (measured on NVIDIA GeForce RTX 2080 Ti) \\
 & -- 66 ms per 10 ms audio frame (measured on 4-core Intel Xeon Gold 6152 CPU, ONNX Runtime) \\
\hline
\end{tabular}
\label{complexity}
\end{table}

\begin{table*}
\caption{Performance comparison of different model setups before and after 8-bit quantization. WF, PF, and EF denote without fine-tuning, partial fine-tuning, and entire fine-tuning, respectively, on AV-HuBERT. Quantization introduces only minor degradation in PESQ, while STOI scores slightly improve across all SSL-based models.}
\centering
\begin{tabular}{|l|c|c|c|c|c|c|}
\hline
 & PESQ(Orig) & PESQ (8-bit) & $\Delta$ PESQ & STOI (Orig) & STOI (8-bit) & $\Delta$ STOI\\  
\hline
\textbf{AVSE} & 1.245 & 1.237 & -0.009 & 0.605 & 0.604 & -0.001\\
\hline
\textbf{SSL-AVSE (WF)} & 1.299 & 1.253 & -0.046 & 0.630 & 0.641 & +0.011\\
\hline
\textbf{SSL-AVSE (EF)} & 1.374 & 1.339 & -0.036 & 0.661 & 0.664 & +0.003\\
\hline
\textbf{SSL-AVSE (PF)} & \textbf{1.396} & \textbf{1.361} & -0.035 & \textbf{0.682} & \textbf{0.686} & +0.004\\
\hline
\end{tabular}
\label{quantization}
\end{table*}

It is important to note that the performance of our system was evaluated using a vocoder-based CI simulation framework. In CI research, vocoder-based CI simulation has become a widely used method, as supported by prior studies \cite{shannon1995speech, dorman1997speech, friesen2001speech}. Although such simulations cannot fully capture the auditory experience of actual CI users, they have been shown to reproduce similar behavioral performance trends. This approach offers several key advantages. First, it reduces variability caused by individual factors in CI users, such as etiology of hearing loss, duration of deafness, and electrode placement. Second, it reduces the fatigue and discomfort that CI users may experience during extended testing sessions, thereby preventing inaccurate or biased results. Third, using vocoded speech with normal-hearing participants allows researchers to isolate and evaluate the effects of signal processing algorithms without confounds from CI hardware differences or user-specific neural adaptation. Given recent advances in CI technology, many studies now use 16-channel tone vocoders \cite{yu2018effects, crew2012channel}. Accordingly, we adopt the same configuration in this study.

To evaluate the feasibility of deploying the proposed SSL-AVSE system in real-world applications, we conducted a detailed analysis of its model complexity and computational efficiency. As summarized in Table \ref{complexity}, the SSL-AVSE system comprises 106 million (M) parameters (103 M for the AV-HuBERT encoder) and requires 27.5 GFLOPs per inference. Profiling results further indicate that the model achieves 5 ms processing time per 10 ms audio frame on the GPU and 66 ms per frame on a 4-core CPU. In addition, we investigated the impact of post-training 8-bit quantization, which reduced the overall model size from 423 MB to 107 MB. As shown in Table \ref{quantization}, quantization introduced only negligible degradation in PESQ, while STOI scores improved slightly across all SSL-based models. We hypothesize that this effect may stem from quantization discarding non-essential variations, thereby sharpening intelligibility-related representations.

These findings collectively demonstrate that the quantized SSL-AVSE system is already well-suited for deployment on more powerful external computing platforms, such as personal computers, smartphones, smart glasses, or tablets, which can readily accommodate the required computational resources. In the context of CI, such devices could act as intermediate processing units to deliver enhanced speech signals to CI sound processors. Existing commercial solutions already support this type of integration. For instance, the Cochlear Wireless PhoneClip streams Bluetooth audio to the Nucleus 6 (CP910) sound processor \cite{huth2022effect}, whereas AudioLink serves as a universal wireless streamer, transmitting audio from external devices (e.g., phones, tablets, TVs, and remote microphones) directly to MED-EL sound processors \cite{muck2023effects}. More recent CI processors, such as the Nucleus 7 (CP1000), are capable of directly interfacing with compatible Apple devices (Made for iPhone, MFi) and Android devices (Android Streaming for Hearing Aids, ASHA) \cite{thibodeau2021advanced}. These established integration pathways suggest that the proposed system could feasibly be incorporated into current CI user ecosystems in the near term.

Looking ahead, further model compression strategies, including pruning and knowledge distillation, are expected to significantly reduce the computational cost and memory footprint of the SSL-AVSE system. These advances would open the possibility of directly deploying the system on CI processors with stringent resource constraints, thereby eliminating the reliance on external edge devices. Ultimately, such developments would extend the benefits of advanced SE technologies to CI users in a wider range of real-world acoustic environments.


\section{Conclusion}
We propose a novel AVSE model, SSL-AVSE, which leverages a pretrained audio-visual foundation model to enhance speech quality and intelligibility in CI scenarios. Although the model was pretrained using an English corpus, it performed well on SE tasks involving Mandarin datasets, demonstrating the ability to generalize to different target languages. Compared with other state-of-the-art AVSE methods, our proposed model resulted in a notable increase in speech quality and word intelligibility. 
Although many previous studies have demonstrated the effectiveness of pretrained models on various downstream tasks, this work is the first to apply a pretrained audio-visual model to enhance SE performance and demonstrate its potential benefits for CI users. We believe that this study represents a promising direction for advancing AVSE technologies across a range of assistive listening devices, including hearing aids, CIs, and personal sound amplification products (PSAPs). Moving forward, we aim to reduce the model size of the proposed SSL-AVSE system through techniques such as parameter pruning and knowledge distillation. As modern devices increasingly incorporate sensors to capture multimodal data, we also plan to explore the integration of additional modalities, such as tactile and textual information, to further improve SE performance. This represents another important avenue for future research.

\bibliographystyle{IEEEbib}
\bibliography{ref.bib}

\end{document}